\documentstyle[preprint,prd,aps]{revtex}
\begin{document}
\draft
\preprint{WUGRAV-95-12}
\title{Comment on the paper of Leonard Parker and Yang Zhang \\
``Cosmological perturbations of a relativistic condensate'' }
\author{L. P. Grishchuk}
\address{McDonnell Center for Space Sciences, Physics Department\\
Washington University, St. Louis, Missouri 63130\\
and\\
Sternberg Astronomical Institute, Moscow University\\
119899 Moscow, V234, Russia\\
(May 12, 1995, Submitted to Physical Review D)}
\maketitle
\begin{abstract}
The ``standard'' inflationary formula for density perturbations is
often being used in the literature and, in particular, it has been used in
the paper of Parker and Zhang. Among other things, this formula suggests
that the contribution of density perturbations to the microwave background
anisotropies is much larger than the contribution of gravitational waves in
the limit of the de Sitter inflation. It is shown that this formula is
incorrect.
\end{abstract}
\vskip 2cm
\pacs{PACS numbers:  98.80.Cq, 04.30.-w, 42.50.Dv, 98.70.Vc}
\newpage
In a recent paper, Parker and Zhang [1] consider cosmological
perturbations that can be possibly produced in the early Universe.
This is an interesting problem having important observational
implications. The paper of Parker and Zhang is transparent
and honest, in the sense that the authors clearly identify
what they derive and what they use from the previous literature.
The final numerical estimates of the paper~[1] rely entirely
on a formula which the authors take from the inflationary
literature. This formula, see Eqs.~(22), (43), relates
the amplitude of perturbations today with the amplitude
of perturbations during the inflationary stage. The formula
suggests an enormous increase of the amplitude, if the equation
of state at the inflationary phase has happened to be sufficiently
close to the de Sitter one. The point of my comment is that
this formula is incorrect, as we will see below, and the results
based on this formula cannot be trusted.

It appears that the use of this formula is connected to a certain
disorientation as for the mechanism responsible for the production
of cosmological perturbations. In the introductory part of their paper,
Parker and Zhang associate the perturbations with the existence
of a particle horizon in de Sitter space. If it is the particle
horizon that is responsible for the generation of perturbations,
then the question arises why the perturbations cannot be generated
by the particle horizon of the radiation-dominated Universe.
Later, Parker and Zhang associate the perturbations with the
``horizon Hawking temperature'' (apparently, they refer
to the paper of Gibbons and Hawking~[2] where the event
horizon of the exact de Sitter solution was considered).
The notion of the event horizon is global, not local.
In realistic cosmological models that are usually discussed,
there is no event horizon at all, despite the possible presence
of an intermediate stage of the quasi-de Sitter expansion.
If it is the event horizon that is responsible for the generation
of cosmological perturbations, then --- there is no event horizon,
there is no horizon temperature, there is nothing to discuss.
As a result, the authors of Ref.~[1] were not surprised
by the formula which essentially states that one can produce
an arbitrarily large amount of density perturbations by
practically doing nothing.

The formula used in [1] has been derived as a continuation
of the previous studies on this subject. Parker and Zhang
refer to the papers~[3-8]. We will start from the paper
of Hawking~[7] which seems to be clearer than others
in expressing the basic idea and intentions.
The papers~[5,6,7] are similar in many respects.

Hawking considers a scalar field $\phi$ running slowly down
an effective scalar field potential. He discusses the
inhomogeneous fluctuations
$\phi_1(t,{\bf x})$
in the field
$\phi=\phi_0(t)+\phi_1(t,{\bf x})$
which mean that on a surface of constant time there will be some
regions where the $\phi$ field has run further down the hill
than in other regions. He introduces a new time coordinate
$\bar{t}=t+\delta t(t,{\bf x})$
in such a way that the variations of the field are removed
and the surfaces of constant time are surfaces of constant $\phi$.
Since the scalar field transforms as\\
$\phi_0+\phi_1 \rightarrow \phi_0+\phi_1 - \dot{\phi}_0 \delta t$,
the required condition is achieved by the time coordinate shift
$\delta t=\phi_1/\dot{\phi}_0$.
Note that for a given $\phi_1$ the time shift is larger,
the smaller is $\dot{\phi}_0$. Then Hawking says that the change
of time coordinate will introduce inhomogeneous fluctuations
in the rate of expansion $H$. He and other authors take
$\delta H\sim H^2\delta t$.
{}From here they come, implicitly or explicitly,
to the dimensionless amplitude of density perturbations
\begin{equation}
  {\delta\rho \over \rho} \sim {\delta H\over H}
\sim H\delta t \sim {H\phi_1 \over \dot{\phi}_0} \quad .
\end{equation}
Some authors write explicitly
$\phi_1 \sim H$
and $\delta\rho / \rho \sim H^2 / \dot{\phi}_0$.

The analysis has been done at the inflationary stage.
To obtain the today's amplitude of density perturbations
in wavelengths, say, of the order of the today's Hubble radius,
it is recommended to calculate the right hand side of Eq.~(1)
at the moments of time when the scales of our interest were
``crossing horizon'' during inflationary epoch. In one or another
version this formula appears in the most of inflationary literature
and because of numerous repetitions it has grown to the "standard" one.
According to this formula, the amplitude of density perturbations
becomes larger if one takes the $\dot{\phi}_0$ smaller.

The authors of [5,6,7] work with a specific scalar field potential,
so the numerical value of $\dot{\phi}_0$ and the numerical value
of $\delta\rho / \rho$ following from Eq.~(1) turn out
to be dependent on the self-coupling constant in the potential.
These authors are concerned about the unacceptably large amplitude
of density perturbations that they have produced.
But it is not a concern about the fact that the Einstein
equations play no role in this argumentation.
It is a tricky detail in the scalar field potential that
the authors of [5,6,7] do not like.

Now let us show what is wrong with the argumentation of~[5,6,7].
Let us consider a scalar field $\phi$ with arbitrary potential.
Write the field as $\phi=\phi_0(t)+\phi_1(t)Q$ where $Q$
is the $n$-th spatial harmonic, $Q^{,i}_{,i}+n^2Q=0$.
Write the perturbed metric in the form
\[
ds^2 = -dt^2 + a^2(t)[(1+h(t)Q)\delta_{ij}
     + h_l(t)n^{-2}\, Q_{,i,j}]dx^i\, dx^j \quad .
\]
The de Sitter solution corresponds to $\dot{\phi}_0 =0$,
$a(t)\sim e^{Ht}$, and
$H(t)=\dot a /a={\rm const}$.
It follows from the Einstein equations that the (linear)
contribution $\epsilon_\phi$ of the scalar field perturbations
to the total energy density $\epsilon =\epsilon_0+\epsilon_\phi$
can be written as
\[
  \epsilon_\phi = \dot{\phi}_0 \left\{ \dot{\phi}_1-\phi_1
\big[ \ln(a^3\dot{\phi}_0)\big]^\cdot\right\} Q \quad .
\]
The contribution $\epsilon_\phi$, as well as other components
of the perturbed energy-momentum tensor, vanish in the de Sitter
limit $\dot{\phi}_0 \rightarrow 0$. Thus, the first conclusion
we have to make is that in the de Sitter limit there is
no linear density perturbations at all. The scalar field
perturbations are uncoupled from gravity, they are not
accompanied by linear perturbations of the energy-momentum
tensor and they are not accompanied by linear perturbations
of the gravitational field. The general solution to the
perturbed Einstein equations is a set of purely coordinate
solutions totally removable by appropriate coordinate
transformations. The scalar field perturbations reduce to a test
field whose role is to identify events in the spacetime.
One can still ask about a coordinate
system such that the surfaces of constant time $\tau$,
$\tau=\phi_1(t,{\bf x})$ are surfaces of constant $\phi$.
But the perturbation of the expansion rate of this new
coordinate system will have nothing to do with the energy
density perturbations. [An attempt of cutting the de Sitter
space-time along a surface of the new time and simply joining
it to the radiation-dominated stage would have shown that
perturbations at the radiation-dominated stage are completely
determined by the coordinate solutions at the de Sitter stage.
This result would have only signaled about a mistake that
has been made. Indeed, the perturbed equations become singular
at the matching surface and dealing with the solutions requires
special care, see Ref.~[12].]

Now let us assume that $\dot{\phi}_0$ is not zero.
Transformation of time $\bar{t}=t+\chi (t)Q$ generates a Lie
transformation of the scalar field:
\[
  \phi_0(t)+\phi_1(t)Q \rightarrow \phi_0(t)
+ [\phi_1(t)-\dot{\phi}_0(t)\chi (t)]Q \quad .
\]
If one wants the transformed field to be homogeneous one takes
$\chi (t)=\phi_1(t)/\dot{\phi}_0(t)$. The same transformation
of time generates Lie transformations of the metric.
The transformed $g_{oo}$ component is
$\bar{g}_{oo} =-1+2\dot{\chi} Q$, the transformed $g_{ik}$
components are described by\\
$\bar{h}=h-2(\dot a /a)\chi$.
There appear also the $g_{oi}$ components but they will not
participate in our linear analysis. The expansion rate of
a given frame of reference is determined by the trace of
the deformation tensor~[9]:
\[
  D = {1\over 2\sqrt{-g_{oo}}}\,
{\partial (g_{ik}-g_{oi}g_{ok}/g_{oo})\over\partial t}\,
  g^{ik} \quad .
\]
In the linear approximation and before the transformation,
\[
  D \approx 3H + {1\over 2} (3\dot h - \dot h_l)Q \quad .
\]
After the transformation,
$\bar{D} = D - 3\dot H \chi Q$.

So, the introduced inhomogeneous fluctuation in the rate
of expansion is $\delta H = -\dot H \chi Q =\dot H \delta t$,
not $\delta H = H^2\delta t$ assumed in Refs.~[5,6,7].

The Einstein equation for energy density
$D^2/3 = \kappa\epsilon$ is satisfied before
and after the transformation, since the variation of $H$
is balanced by the variation of the energy density.
The transformed energy density is
\[
 \bar{\epsilon }
= \epsilon_0 + \epsilon_\phi - \dot\epsilon_0 \chi Q
= \epsilon_0 + \dot{\phi}_0^2
  \left( {\phi_1 \over \dot{\phi}_0}\right)^\cdot Q \quad .
\]
Thus, if one makes the $\dot{\phi}_0$ smaller, the energy
density perturbation decreases according to the Einstein
equations, and it increases according to the conjectures
of Refs.~[5,6,7].

The situation becomes even more disturbing if one recalls
that the formula (1) has been seemingly confirmed by more
detailed studies. People did really write the perturbed
Einstein equations. Moreover, it was done in the framework
of the so-called gauge-invariant formalism, the whole purpose
of which is to eliminate coordinate solutions and to work
exclusively with something ``physical''. Parker and Zhang
quote the paper~[4]. It is useful to consider also the
paper~[10] which summarizes the previous work and gives a
clearer exposition.

In terms of the gauge-invariant potential $\Phi$, the basic
equation of [10] is
\begin{equation}
 \Phi^{\prime\prime} + 2
 {(a/\phi_0^\prime )^\prime \over (a/\phi_0^\prime )}
 \Phi^\prime - \nabla^2 \Phi + 2\phi_0^\prime
\left( { {\cal H} \over \phi_0^\prime}\right)^\prime
       \Phi = 0
\end{equation}
where ${}^\prime = d/d\eta$, $dt=ad\eta$, ${\cal H} = a^\prime /a$,
$\nabla^2\Phi =-n^2\Phi$.
Equation~(2) is exactly the same equation as the basic
equation~(2.23) of Ref.~[4].  These equations were derived
from the original perturbed Einstein equations with the help
of manipulations aimed at expressing the equations in terms
of the gauge-invariant potentials.  Parker and Zhang refer
to a conservation law found in Ref.~[4]. Indeed, in terms
of the quantity $\zeta$ defined as
\[
  \zeta = {2\over 3}\, {H^{-1}\dot\Phi + \Phi \over 1+w}
        + \Phi \quad ,
\]
where $w=p/\epsilon$, Eq. (2) takes on the form
\[
   {3\over 2}\, \dot{\zeta}\, H(1+w) = -{n^2\over a^2} \Phi \quad .
\]
In the long-wavelength limit $n^2\rightarrow 0$, the authors
of Refs.~[10] and [4] neglect the right-hand side of this equation
and arrive at the ``conservation law'':
\begin{equation}
   \zeta \approx {\rm const} \quad .
\end{equation}
They use the constancy of $\zeta$ all the way from the first
``horizon crossing'' at $t_i$ to the second ``horizon crossing''
at $t_f$. Returning to the definition of $\zeta$ and remembering
that $1+w(t_f)$ is of the order of 1 while $1+w(t_i)$ is much
smaller than 1, one can derive
\begin{equation}
  \Phi (t_f) \sim \Phi (t_i) [1+w(t_i)]^{-1} \quad .
\end{equation}
The authors of Refs.~[4,10] emphasize that this formula is
in agreement with Eq.~(1). It is essentialy this formula
that has been used by Parker and Zhang.   Equation~(4)
suggests an arbitrarily large production of density perturbations
for no other reason but simply because the $1+w(t_i)$ was very
close to zero. This formula cannot be correct.
[I realize perfectly well that what I qualify here as obviously
incorrect is definitely considered by others as obviously correct.
Otherwise somebody would raise a voice of protest against
the ease with which inflationists generate tremendous amounts
of various substances (some of them are even claiming that they
can ``overclose'' our Universe). However, judging from the
literature, it is not only that there are no voices of protest
but there is rather an element of competition as for who was
the first to proclaim the ``standard'' inflationary results.
For instance, the authors of~[11] address the inflationary
claims about density perturbations as ``first quantitatively
calculated in~[7,5,6] [and which] have been successfully
quantitatively confirmed by the COBE discovery''.]

Now let us show what is wrong with the derivation of Eqs.~(3)
and (4).  Use the background equations in order to express
the coefficients of Eq.~(2) in terms of the scale factor
$a(\eta )$ and its derivatives. Introduce a new variable
$\mu$ according to the definition
\begin{equation}
  \Phi = {1\over 2n^2}\, {a^\prime \over a} \gamma
  \left( {\mu\over a\sqrt\gamma}\right)^\prime
\end{equation}
where $\gamma = -\dot H/H^2 = 1+(a/a^\prime )^\prime$.
So far, the variable $\mu$ is simply a new variable replacing
$\Phi$, but the importance of $\mu$ is in that the original
perturbed Einstein equations require this variable to satisfy
the equation
\begin{equation}
  \mu^{\prime\prime} + \mu
\left[ n^2 -
       {(a\sqrt\gamma )^{\prime\prime} \over a\sqrt\gamma}
\right] = 0 \quad .
\end{equation}
With the help of Eq. (5), Eq. (2) identically transforms to
\begin{equation}
\Bigg[ \mu^{\prime\prime} + \mu
\left[ n^2 - {(a\sqrt\gamma )^{\prime\prime}
       \over a\sqrt\gamma} \right] \Bigg]^\prime
- {(a\sqrt\gamma )^\prime \over a\sqrt\gamma}
\Bigg[ \mu^{\prime\prime} + \mu
\left[ n^2 - {(a\sqrt\gamma )^{\prime\prime}
       \over a\sqrt\gamma} \right] \Bigg] = 0 \, .
\end{equation}
If Eq. (6) is satisfied, Eq. (7) is satisfied too,
but not {\it vice versa}.  Equation~(7) is equivalent to
\begin{equation}
   {1\over a^2\gamma} \Bigg[ a^2\gamma
    \left( {\mu\over a\sqrt\gamma}\right)^\prime \Bigg]^\prime
   + n^2 {\mu\over a\sqrt\gamma} = X \quad ,
\end{equation}
where $X$ is arbitrary constant. Use the definition of $\zeta$
and Eq.~(5) to show that
\[
\zeta = {1\over 2n^2}\, {1\over a^2\gamma}
\Bigg[ a^2\gamma \left( {\mu\over a\sqrt\gamma}\right)^\prime
\Bigg]^\prime \quad .
\]
In the lowest approximation of $n^2 \rightarrow 0$ the second term
in Eq.~(8) can be neglected. This gives
$\zeta\approx X/2n^2 ={\rm const}$ and explains the
origin of Eq.~(3).

Thus, the ``conservation law'' (3)
can only be used for the derivation of Eq.~(4) if one
is willing to make a mistake, that is to forget that the constant
$X$ must be equal to zero.

The possible relative contributions of density perturbations and
gravitational waves to the observed microwave background anisotropies
is a subject of active study. In the center of discussion are
usually the "consistency relations" which say that the gravity wave
contribution goes to zero if the spectrum of perturbations approaches
the Harrison-Zeldovich form. In reality, these "consistency
relations" are simply a manifistation of inconsistency of the "standard"
inflationary theory from which they are derived.

In conclusion, if the ``standard'' inflationary results are
incorrect and cannot be trusted, what is the amount of density
perturbations that can be generated in the early Universe?
My part of answer is formulated in Ref.~[12].

\end{document}